\documentclass[final,hidelinks,onefignum,onetabnum]{siamart251216}

\usepackage{graphicx}
\usepackage{bm}
\usepackage{amsmath, amssymb}
\usepackage[normalem]{ulem}
\usepackage{lipsum}
\usepackage{appendix}
\usepackage{physics}
\usepackage{xcolor}
\usepackage{appendix}
\usepackage{mathtools}
\usepackage{subfigure}
\usepackage{nicefrac}
\usepackage{comment}
\usepackage{booktabs}
\usepackage{epstopdf}
\usepackage{algorithmic}
\usepackage{hyperref}
\usepackage[all]{hypcap}
\usepackage{cite}

\newcommand{\QFT}[0]{\mathrm{QFT}}

\newcommand{\figref}[2]{\hyperref[#1]{\autoref*{#1}#2}}
\newcommand{\aref}[1]{\hyperref[#1]{App.~\ref*{#1}}}

\ifpdf
  \DeclareGraphicsExtensions{.eps,.pdf,.png,.jpg}
\else
  \DeclareGraphicsExtensions{.eps}
\fi

\newcommand{\waveop}{K}

\newsiamremark{remark}{Remark}
\newsiamremark{hypothesis}{Hypothesis}
\crefname{hypothesis}{Hypothesis}{Hypotheses}
\newsiamthm{claim}{Claim}
\newsiamremark{fact}{Fact}
\crefname{fact}{Fact}{Facts}

\headers{
Quantum Signal Processing: Differential Equations
}{H. Kim, R. Jambunathan, J. Balewski, D. Camps}

\title{
Quantum Signal Processing for Linear PDEs:\\ Circuit Design and Experimental Validation
\thanks{Submitted to the editors \today.
\funding{H.K. was partially supported by grants from the U.S. National Science Foundation under Grant No. DMR-241254 and the U.S. Air Force Office of Scientific Research (AFOSR) under Grant No. FA9550-21-1-0342. This research was supported by and used resources of the National Energy Research Scientific Computing Center (NERSC), a U.S. Department of Energy Office of Science User Facility located at Lawrence Berkeley National Laboratory, operated under Contract No. DE-AC02-05CH11231}.}}

\author{Hyeongjin Kim\textsuperscript{\dag,\ddag}\hspace{-0.45em}
\and Revathi Jambunathan\footnotemark[3]\hspace{-0.2em}
\and Jan Balewski\footnotemark[3]\hspace{-0.2em}
\and Daan Camps\footnotemark[3] 
}
\begin{NoHyper}
\begingroup
  
  \footnotetext{Boston University, Boston, MA (\email{hkim12@bu.edu}).}
  
  \footnotetext{\raggedright Lawrence Berkeley National Laboratory, Berkeley, CA (\email{rjambunathan@lbl.gov}, \email{balewski@lbl.gov}, \email{daancamps@gmail.com}).}
\endgroup
\end{NoHyper}

\usepackage{amsopn}

\ifpdf
\hypersetup{
  pdftitle={Quantum Signal Processing for Linear PDEs: Circuit Design and Experimental Validation},
  pdfauthor={H. Kim, R. Jambunathan, J. Balewski, D. Camps}
}
\fi

\begin{document}

\maketitle

\begin{abstract}
Quantum algorithms offer new avenues for solving partial differential equations (PDEs). While the potential for end-to-end quantum advantage is at present not well understood, recent literature presents explicit circuit constructions for solving certain classes of linear PDEs in the frequency domain and thus offers concrete examples to study. In this work, we develop end-to-end implementations of these quantum circuits compiled to machine-level instructions and benchmark them in both numerical simulations and IBMQ hardware experiments. We focus on the advection, wave, and Poisson equations and study quantum circuits that propagate the dynamics in frequency space via the quantum Fourier transform using approximate methods based on a first-order approximation which offer compact representations with uncontrollable approximation error, and polynomial approximation methods based on quantum signal processing (QSP) leading to deeper circuits with tunable algorithmic error. In addition, we experimentally demonstrate that the QSP-augmented algorithm can provide accurate solutions under realistic hardware constraints. Finally, we extend our method to address non-homogeneous Dirichlet boundary conditions and verify it numerically for a Poisson equation with source term obtained from high-fidelity physics simulations of a capacitively coupled plasma.
\end{abstract}



\section{Introduction}

Quantum computers~\cite{Nielsen2010} are expected to provide computational advantage over classical computers for select computational tasks~\cite{shor1994,grover1996,harrow09} (see Ref.~\cite{ladd2010quantum,montanaro2016quantum,grigoryan2025quantum,camps2025quantumcomputingtechnologyroadmaps} for reviews). However, current quantum hardware capabilities are severely limited by scale and noise. Scaling up quantum computers will ultimately depend on fault-tolerance and on using many physical qubits to encode fewer logical qubits~\cite{eisert2025mindgapsfraughtroad}. The noisy intermediate-scale quantum (NISQ) regime~\cite{preskill2018nisq} uses physical qubits directly for logical computation and relies on error mitigation strategies that improve the signal-to-noise ratio in the computation at the cost of additional circuit samples. NISQ devices have been extensively used to develop and test quantum applications in realistic settings and have demonstrated the capability to deliver quantum advantage on highly-specialized tasks~\cite{google2019quantumsupremacy,quantinuum2025supremacy}.

One proposed application of quantum computing is solving partial differential equations (PDEs)~\cite{evans10}. PDEs are central to modeling and predicting phenomena across science and engineering, from turbulent transport in fluids, to molecular interactions in quantum chemistry, to price dynamics in quantitative finance. Many of these problems require simulations spanning multiple orders of magnitude in resolution, requiring very fine grids to achieve accurate results and driving the need for large-scale simulations. Classical computing, constrained by the approaching limits of Moore’s Law and the growing demands of these simulations, is increasingly strained. Quantum computing promises a path to overcome some of these limitations by offering an exponential advantage in memory efficiency, where an $n$-qubit system can represent $N=2^n$ grid points. However, state preparation and readout remain challenging on quantum hardware~\cite{Aaronson2015}. 

Variational quantum algorithms have been proposed for linear and nonlinear PDEs~\cite{lubasch19,kryiienko2021,jaksch2023variational,pool24}. These approaches typically employ a hybrid classical-quantum workflow that translates the system of differential equations into a minimization problem, where the loss function is evaluated using a quantum computer and variationally minimized using a classical computer. However, these approaches are typically devoid of theoretical convergence or validity guarantees due to their variational nature~\cite{wang2021noise,larocca2025barren}. To obtain theoretical guarantees, some works explicitly construct quantum circuits for solving linear PDEs. In particular, many successful approaches translate the differential equations into an equivalent problem in which quantum algorithms exist or can be explicitly constructed. Common examples include converting PDEs to linear systems of equations~\cite{Berry_2014,Childs2020,Childs2021,lloyd2020,liu21,tennie2024} (\textit{linearization}) or to equivalent Hamiltonian simulation problems~\cite{Shi24,hu2024quantum,jin2025quantum} (\textit{Schrödingerization}).

Another approach, based on spectral methods, transforms the system of differential equations to Fourier space using the quantum Fourier transform (QFT)~\cite{coppersmith2002approximatefouriertransformuseful,Nielsen2010}. This leads to efficient and explicit quantum circuit designs for linear differential equations~\cite{wright2024noisy,lubasch2025}. In this work, we focus on three different linear PDEs: the advection equation, the wave equation, and the Poisson equation. For each type, we explicitly construct quantum circuits using QFT to solve the equation either by propagating the advection or wave dynamics in time, or by implementing the (pseudo)inverse of the Laplacian for the Poisson equation. QFT provides access to diagonal representations of translation-invariant operators, enabling compact implementations of differential operators in Fourier space. Quantum signal processing (QSP)~\cite{gilyen,martynGrandUnificationQuantum2021,motlagh2024generalized, yamamotoRobustAngleFinding2024} allows for explicit circuit constructions with tunable error for implementing these diagonalized differential operators, while a small-angle approximation (SAA) based on a first-order approximation offers an alternative compact implementation with uncontrollable error. QSP has previously been used in the literature to solve differential equations~\cite{lubasch2025,novikau2025efficientexplicit,novikau2025quantumalgorithmadvection}.

\begin{figure}[!t]
    \centering
    \includegraphics[width=\columnwidth]{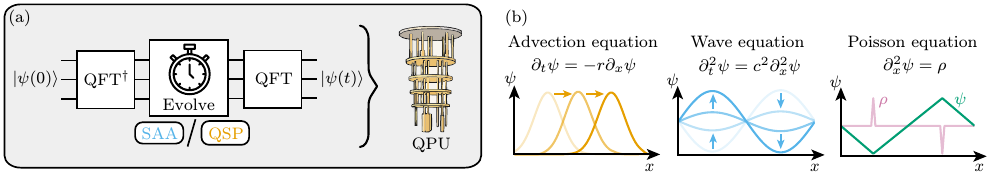}
    \caption[]{(a) Our methodology for running the quantum circuit on a quantum processing unit (QPU) consists of (1) an initial state preparation step to load the state $\ket{\psi(x,t=0)}$ at time $t=0$, (2) an inverse quantum Fourier transform (QFT) to transform the problem to frequency space, (3) solve the system in frequency space using small-angle approximation (SAA) or quantum signal processing (QSP), and (4) transforming back to position space using QFT to obtain $\ket{\psi(x,t)}$. This prepares the state $\ket{\psi(x,t)}$ at time $t$. (b) We develop end-to-end implementations for simulating the advection equation (\textbf{left}), wave equation (\textbf{middle}), and Poisson equation (\textbf{right}).
    \label{fig:main}}
\end{figure} 

Our contributions are three-fold: (1) we develop \emph{end-to-end implementations of these quantum algorithms}, from initial state preparation to final solution, as quantum circuits that can be compiled to machine-level instructions and benchmark them via numerical simulations and experimental runs on IBMQ hardware, (2) we experimentally demonstrate that the QSP-based method can provide more accurate solutions than the SAA-based one under NISQ hardware constraints, and (3) we extend the methodology to non-homogeneous Dirichlet boundary conditions, which enables us to quantum simulate the solution for the Poisson equation for a high-fidelity physics simulation of a capacitively coupled plasma. The implementations are made publicly available via a Python package \texttt{qsp4pde}~\cite{qsp4pde}. The quantum simulation workflow is summarized in~\Cref{fig:main}.

\section{Setup} \label{sec:setup}

We lay out the ingredients used to solve the PDEs. As in Ref.~\cite{lubasch2025}, we utilize the quantum Fourier transform (QFT) to represent PDEs in Fourier space. Most of the details can be found in Ref.~\cite{lubasch2025}, but we summarize the methodology here for completeness; our main contributions are the end-to-end circuit implementations and corresponding numerical and hardware experiments with these algorithms.

\subsection{Discretization and encoding} \label{subsec:background}

We focus on one-dimensional problems. Suppose we have a continuous function $\psi(x)$ on the interval $(-\nicefrac{1}{2}, \allowbreak \nicefrac{1}{2})$ with periodic boundary conditions. The function is discretized and normalized to be encoded in an $n$-qubit wave function $\ket{\psi}$ representing values on a grid of $N=2^n$ symmetrically distributed points around zero. The grid points are defined as
\begin{equation}
\begin{split}
x & \in \{-\nicefrac{1}{2}+\nicefrac{1}{2N}, -\nicefrac{1}{2}+\nicefrac{3}{2N}, \ldots, \nicefrac{1}{2}-\nicefrac{3}{2N}, \nicefrac{1}{2}-\nicefrac{1}{2N} \} \\
  & = \{x_0, x_1, \hdots, x_{N-1}\} \quad \text{where} \quad 
    \begin{aligned}[t]
       x_j &= -\nicefrac{1}{2} + \nicefrac{1}{2N} + \nicefrac{j}{N} \\
           &= \nicefrac{j}{N} - \nicefrac{N-1}{2N},\, j \in [N]
    \end{aligned}
\end{split}
\label{eq:xgrid}    
\end{equation}
Each point $x_j$ is matched with the computational basis state $\ket{j}\!=\!\ket{\texttt{j}_0,\!\texttt{j}_1,\!\ldots,\!\texttt{j}_{n-1}}$, where $\texttt{j}_q \in \{0,1\}$ corresponds to the state of qubit $q$ and $j = \sum_{q=0}^{n-1} 2^q \texttt{j}_q$ follows the little-endian convention. The function value $\psi(x_j)$ is then encoded in the amplitude of the wave function on the state $\ket{j}$, i.e., $\braket{j}{\psi} = \psi(x_j) / \norm{\psi}$, where $\norm{\psi} \equiv \sqrt{\sum_{j} \abs{\psi(x_j)}^2}$.

\subsection{Fourier representation of differential operators}

A crucial ingredient for solving the linear PDEs is the quantum Fourier transform (QFT)~\cite{coppersmith2002approximatefouriertransformuseful,Nielsen2010,camps21}, which converts from the position variable $x_j$ to the reciprocal space where the spatial frequency is denoted by the wavenumber $k$ (and vice versa). Since the spatial coordinates are centered around zero, we use the \emph{shifted} QFT~\cite{Nielsen2010}. In our implementation of the QFT, the Fourier basis states are $\ket{k} = \ket{\texttt{k}_0, \texttt{k}_1, \ldots, \texttt{k}_{n-1}}$, where $\texttt{k}_q \in \{0, 1\}$ and $k =\sum^{n-1}_{q = 0} 2^{n-1-q} \texttt{k}_q$ (big-endian). The shifted QFT maps the Fourier wavenumber coordinates $k$ to the spatial coordinates $x_j$; in other words, the input states of the QFT are wavenumbers $\ket{k}$ while the output states are coordinates $\ket{j}$:
\begin{align} \label{eq:QFT}
\begin{split}
    \mel{j}{\QFT}{k} &= \frac{1}{\sqrt{N}} e^{{\mathrm i} 2\pi\left(k - \frac{N}{2} \right) \left(\frac{j}{N}-\frac{N-1}{2N}\right)} = \frac{1}{\sqrt{N}} e^{{\mathrm i} 2\pi \tilde{k} x_j},
\end{split}
\end{align}
with shifted wavenumbers,
\begin{equation}
\tilde{k}  = k - \nicefrac{N}{2} \in \{-\nicefrac{N}{2}, -\nicefrac{N}{2}+1, \ldots, \nicefrac{N}{2}-1\},    
\label{eq:wavenumbers}
\end{equation}
and spatial coordinates $x_j$ as defined in \Cref{eq:xgrid}. The shifted QFT has an efficient quantum circuit representation with $\tfrac{n(n-1)}{2}$ controlled two-qubit rotation gates, $n$ Hadamard gates, and $2n$ single-qubit rotation gates (see Ref.~\cite{Nielsen2010}).

The power of QFT comes when dealing with problems that satisfy periodic boundary conditions, where spatial differential operators become diagonal in the reciprocal Fourier basis. For a central finite difference approximation to $\pdv{x}$ (with $\delta_x$ denoting the finite difference derivative) we have that
\begin{equation*}
\begin{split}
    \left.\pdv{x} \mel{j}{\QFT}{k}\right|_{x=x_j} &\approx \delta_x \mel{j}{\QFT}{k}(x_j) =
    \frac{1}{\sqrt{N}} \frac{e^{{\mathrm i}2\pi \tilde{k} x_{j+1}} - e^{{\mathrm i}2\pi \tilde{k} x_{j-1}}}{x_{j+1} - x_{j-1}},
\end{split}
\end{equation*}
which, using $x_{j+1} - x_{j-1} = \nicefrac{2}{N}$, we can write as
\begin{equation}
\begin{split}
    \delta_x \mel{j}{\QFT}{k}(x_j)
    &= N \left(\frac{e^{2{\mathrm i}\pi \tilde{k}/N} - e^{-{\mathrm i}2\pi \tilde{k}/N}}{2} \right) \frac{1}{\sqrt{N}} e^{{\mathrm i}2\pi \tilde{k} x_j} \\
    &= {\mathrm i} N \sin(\frac{2\pi \tilde{k}}{N}) \mel{j}{\QFT}{k}.
\end{split}
\label{eq:cfd}
\end{equation}

We introduce a diagonal operator $\waveop$ to represent the wavenumber $\tilde{k}$ satisfying $\waveop \ket{k} = \tilde{k} \ket{k}$. One can directly verify that
\begin{equation} \label{eq:wavenumber_operator}
    \waveop = -\frac{N}{4} \sum^{n-1}_{j = 0} 2^{-j} Z_j - \frac{1}{2} I,
\end{equation}
where $Z_j$ is the Pauli-$Z$ operator $\left( \begin{smallmatrix} 1 &  \\  & -1 \end{smallmatrix} \right)$ on qubit $j$ and $I$ is the $n$-qubit identity, satisfies $\waveop \ket{k} = \tilde{k} \ket{k}$ following the big-endian convention. Using \Cref{eq:cfd}, the finite difference operator $D_x$ becomes
\begin{equation}
    D_x= {\mathrm i} N \cdot \QFT \sin(\frac{2\pi \waveop}{N}) \QFT^\dagger,
\label{eq:d_op}
\end{equation}
which is diagonal in the Fourier basis because $\waveop$ is diagonal. Following a similar argument for the central difference approximation to the second derivative, $\delta_x^2 \mel{j}{\QFT}{k}\allowbreak(x_j)$, we get that the Laplacian operator $\Delta_x$ is also diagonal in the Fourier basis,
\begin{equation}
    \Delta_x = -4N^2 \cdot \QFT \sin^2{\left(\frac{\pi \waveop}{N} \right)} \QFT^\dagger.
\label{eq:laplacian}
\end{equation} 
Henceforth, $D_x$ and $\Delta_x$ denote the discrete first-difference and discrete Laplacian operators, respectively, which are implemented by the quantum circuits.

\subsection{Laurent-polynomial quantum signal processing}\label{subsec:QSP}

The final quantum algorithmic ingredient that we need is a methodology to implement Laurent-\hspace{0pt}polynomial transformations of a unitary operator. We use Laurent-polynomial quantum signal processing (QSP), specifically the generalized QSP construction of Refs.~\cite{motlagh2024generalized,yamamotoRobustAngleFinding2024,laneve2025,silva2022}, which acts directly on a unitary $U=e^{{\mathrm i}H}$. This setting is closely related to the quantum singular value transform (QSVT)~\cite{gilyen,martynGrandUnificationQuantum2021}, but our constructions do not use block encodings or reflection operators and do not rely on the cosine-sine decomposition. We therefore use QSP terminology throughout this manuscript.

\begin{figure*}[!t]
    \centering
    \includegraphics[width=\columnwidth]{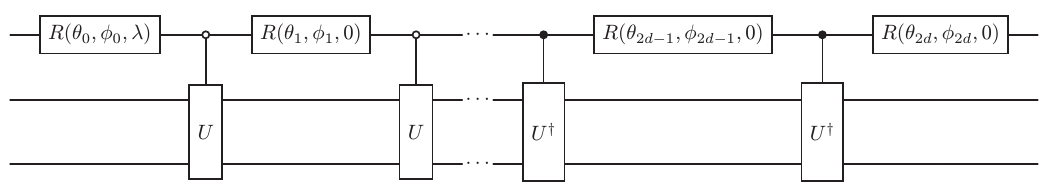}
    \caption[]{Circuit representation of Laurent-polynomial QSP. This circuit is based on \Cref{eq:GQSP} and originates from Ref.~\cite{motlagh2024generalized}. The top wire is for the ancilla qubit, while the rest of the wires are for the $n$ system qubits.
    \label{fig:qsvt}}
\end{figure*} 

Given a Hermitian operator $H = H^{\dagger}$ and the corresponding unitary $U=e^{{\mathrm i}H}$, Laurent-polynomial QSP implements $P(U)$, where
\begin{equation}
    P(z) = \sum^{d}_{m=-d} a_m z^m \in \mathbb{C}[z,z^{-1}]
\end{equation}
is a complex-valued Laurent polynomial with coefficients $a_m \in \mathbb{C}$ such that $\abs{P(e^{{\mathrm i}\varphi})} \leq 1$ for $\varphi \in \mathbb{R}$. To implement $P(U)$, we introduce an ancillary qubit and define the $\mathrm{SU}(2)$ rotation gate (up to a global phase) acting on the ancilla as:
\begin{equation}
    R(\theta, \phi, \lambda) = \begin{bmatrix}
        e^{{\mathrm i}(\lambda + \phi)} \cos(\theta) & e^{{\mathrm i}\phi} \sin(\theta) \\
        e^{{\mathrm i}\lambda} \sin(\theta) & - \cos(\theta)
    \end{bmatrix} \otimes I^{\otimes n}.
\label{eq:su2}
\end{equation}
Then, the generalized QSP theorem~\cite{motlagh2024generalized} states that there exist rotation parameters $\vec{\phi}, \vec{\theta} \in \mathbb{R}^{2d+1}$ and $\lambda \in \mathbb{R}$ such that
\begin{align} \label{eq:GQSP}
    \begin{split}
    \begin{bmatrix}
        P(U) & \cdot \\
        Q(U) & \cdot 
    \end{bmatrix} &= \left(\prod^d_{j=1} R(\theta_{d+j}, \phi_{d+j}, 0) A' \right)
    \cdot \left(\prod^d_{j=1} R(\theta_{j}, \phi_{j}, 0) A \right) R(\theta_0, \phi_0, \lambda) ,
    \end{split}
\end{align}
where 
\begin{equation}
\begin{split}
    A &= (\dyad{0} \otimes U) + (\dyad{1} \otimes I), \\
    A' &= (\dyad{0} \otimes I) + (\dyad{1} \otimes U^\dagger),
\end{split}
\end{equation}
where the products are ordered from left to right as written, and
\begin{equation}
    Q(z) = \sum^{d}_{m=-d} b_m z^m \in \mathbb{C}[z,z^{-1}]
\end{equation}
is a complementary Laurent polynomial with $\abs{P(e^{{\mathrm i}\varphi})}^2 + \abs{Q(e^{{\mathrm i}\varphi})}^2 = 1$ for $\varphi \in \mathbb{R}$. Finding such a $Q$ is not a trivial task. To do this, we implement the robust root-finding method over Laurent polynomials detailed in Ref.~\cite{yamamotoRobustAngleFinding2024}. Empirically, we found that the numerical root-finding method can fail to find a $Q$ that satisfies the normalization for all $\varphi \in \mathbb{R}$. To stabilize the procedure, we rescale $P \to \gamma P$ with $0< \gamma < 1$ before constructing $Q$. This avoids numerical issues, but the rescaling reduces the QSP success probability (i.e., the probability of the ancilla qubit being in the state $\ket{0}$) by a factor of $\gamma^2$.

Once $P(\cdot)$ and $Q(\cdot)$ are fixed, the rotation parameters ($\vec{\theta}$, $\vec{\phi}$, $\lambda$) for the SU(2)-gates in~\Cref{eq:su2} are computed using the matrix-peeling pseudocode from Ref.~\cite{motlagh2024generalized}. In essence, we construct a $2 \times (2d+1)$ matrix $S$ whose first row contains coefficients $a_m$ of $P(\cdot)$ and second row contains coefficients $b_m$ of $Q(\cdot)$. The algorithm iterates from degree $2d$ down to $0$. At each iteration $i$, the angles are extracted from the highest-order coefficients, i.e., $\theta_i = \arctan(\abs{b_i}/\abs{a_i})$ and $\phi_i = \arg(a_i/b_i)$. The corresponding SU(2)-gate $R$ is applied to $S_i$ to iterate to the lower-degree term, $S_{i+1} \gets R S_i$.

Using \Cref{eq:GQSP}, the operator $P(U)$ can be applied to the refrence state $\ket{\psi_0}$ by measuring (or post-selecting) the ancilla qubit in state $\ket{0}$. If we initialize our quantum state as $\ket{0} \otimes \ket{\psi_0}$, where $\ket{0}$ is the state of the ancilla qubit and $\ket{\psi_0}$ is the initial wave function of the system, then
\begin{align} \label{eq:QSP_on_initial_state}
\begin{split}
    \begin{bmatrix}
        P(U) & \cdot \\
        Q(U) & \cdot 
    \end{bmatrix} (\ket{0} \otimes \ket{\psi_0}) 
    &= \ket{0} \otimes P(U)\ket{\psi_0} + \ket{1} \otimes Q(U)\ket{\psi_0} \\
    &= \ket{0} \otimes \alpha \ket{\psi} + \ket{1} \otimes \beta \ket{\zeta},
\end{split}
\end{align}
where $\ket{\psi} \propto P(U) \ket{\psi_0}$ and $\ket{\zeta} \propto Q(U)\ket{\psi_0}$ are normalized to unity and retain phase information, $\alpha^2 + \beta^2=1$, and $\alpha,\beta \in \mathbb{R}^{\geq 0}$. The desired solution lies in the $\ket{0} \otimes \alpha \ket{\psi}$ term, so we track the normalization factor $\alpha = \| P(U) \ket{\psi_0}\|$ to extract the unnormalized solution $\psi$. If $P$ is rescaled as $P \to \gamma P$ during the construction of $Q$, then the implemented post-selected branch is proportional to $\gamma P(U)\ket{\psi_0}$, and the associated success probability depends on both $\gamma$ and the norm of $P(U)\ket{\psi_0}$.

In the PDE constructions below, we use Laurent-polynomial QSP to implement $P(U)$ for diagonal unitary operators in Fourier space. Our numerical implementations of this procedure, combining Refs.~\cite{yamamotoRobustAngleFinding2024,motlagh2024generalized}, are available in \texttt{qsp4pde}~\cite{qsp4pde}.

\subsection{Overview}

In the remainder of the paper, we consider advection (Section~\ref{sec:advection}), wave (Section~\ref{sec:wave}), and Poisson equations (Section~\ref{sec:poisson}). The solution strategy follows the workflow in \figref{fig:main}{(a)}:
\begin{equation} \label{eq:general_solution}
    \ket{\psi(x,t)} = \QFT \cdot M_\mathrm{evolve} \cdot \QFT^\dagger \ket{\psi(x,t=0)},
\end{equation}
where $M_\mathrm{evolve}$ is an operation (implemented as a quantum circuit) that solves the desired PDE in frequency space. $M_\mathrm{evolve}$ is not always unitary (e.g., Poisson), so we use QSP to implement non-unitary operations.

\paragraph{Implementation conventions}
Our circuit implementations follow \figref{fig:main}{(a)}.
For time-dependent problems, the input state is prepared as $\ket{\psi_0}\propto\sum_j \psi(x_j,t=0)\ket{j}$ using \texttt{StatePreparation} in \texttt{Qiskit}; for Poisson, $\ket{\psi_0}=\ket{\rho}$. When QSP is used, post-selected ancilla-$0$ output prepares a state proportional to the desired unnormalized solution, with the scalar factor recovered from the post-selection probability as in \Cref{eq:QSP_on_initial_state}. In numerical simulations, we access the full statevector; in hardware runs, we compare measured probabilities $\abs{\braket{x_j}{\psi}}^2$, which loses the overall sign.

\section{Advection equation}\label{sec:advection}

The one-dimensional advection equation describes the transport of a substance within a fluid and is given by
\begin{equation}\label{eq:advection_equation}
    \pdv{\psi}{t} = -r \pdv{\psi}{x},
\end{equation}
where $r$ is the scalar bulk speed. We discretize the spatial coordinate using a central finite difference scheme as described in~\Cref{sec:setup} to get
\begin{equation}
    \pdv{\psi(x_j)}{t} = -r \delta_x \psi(x_j),
\end{equation}
which in the operator representation becomes,
\begin{equation}
\begin{split}
    \pdv{\ket{\psi}}{t} = -r D_x \ket{\psi}
    = -{\mathrm i} Nr \, \QFT \cdot  \sin(\frac{2\pi \waveop}{N}) \cdot \QFT^\dagger \ket{\psi}.
\end{split}
\end{equation}
This can be viewed as a time-independent Schr\"odinger equation with Hamiltonian $Nr \, \QFT \cdot  \sin(\frac{2\pi \waveop}{N}) \cdot \QFT^\dagger$.
The solution $\ket{\psi(t)}$ at time $t$, for a given initial state $\ket{\psi(t=0)}$, is
\begin{equation} \label{eq:advection_solution}
    \ket{\psi(t)} = \QFT \cdot e^{-{\mathrm i}tNr \sin(2\pi \waveop / N)} \cdot \QFT^\dagger \ket{\psi(t=0)}.
\end{equation}
We describe two circuit implementations of \Cref{eq:advection_solution}.

\paragraph{Small-angle approximation}

The circuit construction requires implementing the diagonal unitary evolution operator $\exp\bigl(-{\mathrm i}tNr$ $\sin{(2\pi \waveop / N)}\bigr)$. A compact approximate method truncates the Taylor series of the sinusoid to first order, $\sin{(2\pi \waveop/N)} \approx 2 \pi \waveop/N$ (most accurate near $\tilde{k} \approx 0$), leading to the \emph{small-angle approximation} (SAA):
\begin{equation}
    \ket{\psi(t)} \approx \QFT \cdot e^{-{\mathrm i} 2\pi t r\waveop} \cdot \QFT^\dagger \ket{\psi(t=0)},
\end{equation}
where
\begin{equation} \label{eq:advection_solution_SAA}
    e^{-{\mathrm i} 2 \pi t r \waveop} = e^{{\mathrm i}\pi t r} \prod^{n-1}_{j=0} e^{{\mathrm i}\pi t r 2^{-j + n - 1} Z_j}.
\end{equation}
We remark that~\Cref{eq:advection_solution_SAA} is a product of single-qubit Pauli-$Z$ rotations and admits a depth-1 quantum circuit. The approximation error is small only for Fourier components satisfying $\abs*{2\pi \tilde{k}/N} \ll 1$, and therefore depends on the Fourier support of the initial state as well as on the evolution time $t$.

\paragraph{Laurent-polynomial QSP}

Alternatively, we implement $\exp\bigl(-{\mathrm i}tNr$ $\sin{(2\pi \waveop / N)}\bigr)$ using a Laurent polynomial series and apply QSP. One approach uses the discrete Fourier transform:
\begin{equation} \label{eq:advection_solution_qsvt}
    e^{-{\mathrm i}tNr \sin(2\pi \waveop / N)} = \sum\limits^{N/2-1}_{m=-N/2} c_m e^{{\mathrm i}2\pi m \waveop/N},
\end{equation}
with coefficients
\begin{equation}
    c_m = \frac{1}{N} \sum^{N/2-1}_{\tilde{k}=-N/2} e^{-{\mathrm i}tNr \sin(2\pi\tilde{k}/N)} e^{-{\mathrm i}2\pi m\tilde{k}/N}.
\end{equation}
To implement \Cref{eq:advection_solution_qsvt}, we use QSP as outlined in \Cref{subsec:QSP} with $U = e^{{\mathrm i}2\pi \waveop/N}$, a unitary that is efficiently implementable in a quantum circuit following~\Cref{eq:advection_solution_SAA}, and a Laurent polynomial of degree $d = N/2$. This is an exact representation on the $N$-point grid, but the degree grows as $N/2$, which may lead to deep QSP circuits for high-precision discretizations.

We can also use an approximate Laurent series based on the Jacobi-Anger expansion:
\begin{equation} \label{eq:advection_solution_QSP_JA}
    e^{-{\mathrm i}tNr \sin(2\pi \waveop / N)} \approx \sum^d_{m=-d} J_m(-tNr) e^{{\mathrm i}2\pi m \waveop/N},
\end{equation}
where $J_m$ is a Bessel function of the first kind. The degree $d$ sets the accuracy of this approximation, which we explore numerically in the next section.

\begin{figure}[!t]
    \centering
    \includegraphics[width=\columnwidth]{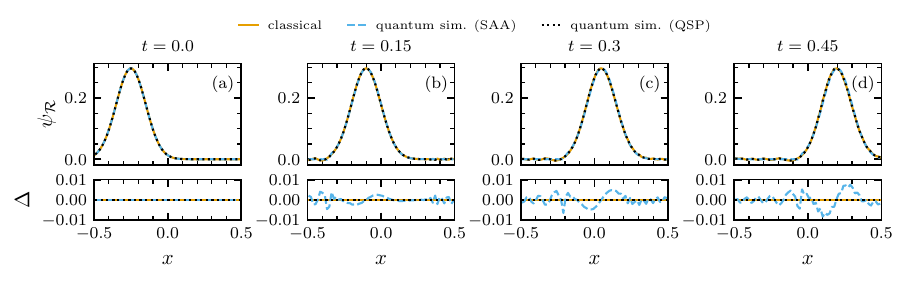}
    \caption[]{Three numerical solutions to the advection equation~\eqref{eq:advection_equation} with $r=1$ and discretized on $2^6 = 64$ grid points. We solve the advection equation starting with the Gaussian wave-packet~\eqref{eq:gaussian_initial}  at times $t = 0, 0.15, 0.3, 0.45$. The real part of the solution $\psi_\mathcal{R}$ is shown on the top panel. Orange solid lines are obtained from a traditional PDE solver $\psi_\mathcal{R}^\mathrm{classical}$. Blue dashed lines and black dotted lines correspond to the quantum circuit solutions using SAA and QSP, respectively. The bottom panel depicts the difference from the classical solution: $\Delta = \psi_\mathcal{R} - \psi_\mathcal{R}^\mathrm{classical}$.
    \label{fig:advection_simulation_2x2}}
\end{figure}

\subsection{Numerical results}\label{sec:adv_numerics}
We solve the advection equation~\eqref{eq:advection_equation} using a shifted Gaussian wave-packet as the initial function:
\begin{equation} \label{eq:gaussian_initial}
    \psi(x,t=0) = \frac{1}{\sigma \sqrt{2 \pi}} e^{-(x-\mu)^2/(2\sigma^2)},
\end{equation}
where $\mu = -0.25$ and $\sigma=0.1$. In \Cref{fig:advection_simulation_2x2}, we compute $\psi(x,t)$ for $t=0,0.15,0.3,0.45$ classically (traditional PDE solver) and with quantum circuits using SAA (from \Cref{eq:advection_solution}) and QSP (with Laurent polynomial derived from the Fourier series in \Cref{eq:advection_solution_qsvt}). Quantum circuits are implemented in \texttt{Qiskit} and simulated with statevectors. We plot the real part of $\psi$, denoted as $\psi_\mathcal{R}$, against the spatial variable $x$. The domain $(-0.5, 0.5)$ is discretized using $2^6 = 64$ grid points following~\Cref{eq:xgrid}, so the quantum simulation uses $n = 6$ qubits to encode $\ket{\psi(t)}$. Bottom panels depict the difference between the quantum and classical results, $\Delta \equiv \psi_\mathcal{R} - \psi_\mathcal{R}^{\mathrm{classical}}$. The QSP solution agrees with the exact classical solution, while the SAA solution accumulates larger error (up to $\abs{\Delta} = 0.01$) as $t$ increases. However, QSP uses more entangling gates than SAA, which is a trade-off that we evaluate further in the hardware experiments (\Cref{sec:hardware}).

\begin{figure}[!t]
    \centering
    \includegraphics[width=\columnwidth]{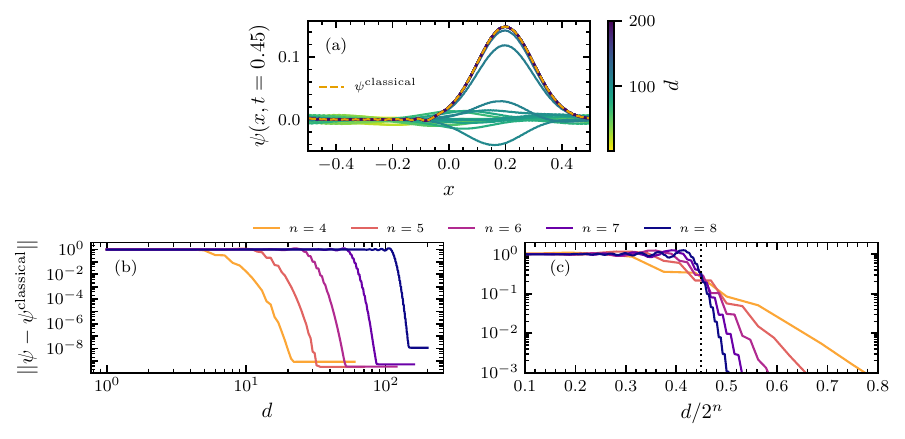}
    \caption[]{Solving the advection equation~\eqref{eq:advection_equation} using the Jacobi-Anger expansion~\eqref{eq:advection_solution_QSP_JA}. We initialize with the Gaussian wavepacket ~\eqref{eq:gaussian_initial} and compute $\psi(x,t=0.45)$. (a) We compare the quantum solution $\psi(x,t=0.45)$ using different truncation orders $d$ against the classical solution $\psi^\mathrm{classical}(x,t=0.45)$ (dashed orange line) for $n = 8$ qubits ($N = 2^8 = 256$ grid points). (b,c) We plot $\norm{\psi - \psi^\mathrm{classical}}$ against $d$ and $d/2^n=d/N$, respectively, for different $n$. The dotted black line indicates the scaling crossover $d^\ast/2^n = 0.45$, aligning with $t = 0.45$.\label{fig:QSP_JA}}
\end{figure} 

Next, we evaluate the convergence of the Laurent-series approximation obtained via the Jacobi-Anger expansion, see \Cref{eq:advection_solution_QSP_JA}. In \figref{fig:QSP_JA}{(a)}, we solve the advection equation at $t = 0.45$ with the Gaussian wavepacket. We set $n=8$ ($256$ grid points) and plot $\psi(x,t=0.45)$ for different truncation degrees $d$. The orange dashed line is the classical ground truth $\psi^\mathrm{classical}(x,t=0.45)$. As $d$ increases, the quantum simulation converges to the ground truth. This is clearer in \figref{fig:QSP_JA}{(b)}, where we plot $\norm{\psi-\psi^\mathrm{classical}}$ versus $d$ for different $n$. The error decreases rapidly with $d$ after the onset of convergence. In \figref{fig:QSP_JA}{(c)},  we plot the same quantity against $d/N$ (where $N = 2^n$), which shows a scaling crossover at $d^\ast/N \approx 0.45$. In particular, it is known that $d$ must scale as $\Theta(tNr)$ (see Refs.~\cite{low2017optimal,low2019hamiltonian}) and thus we numerically observe that $d^\ast \approx 0.45 N$ since $r = 1$ and $t = 0.45$ We use this scaling to vary $d$ when solving the advection equation at different $t$ on IBMQ (\Cref{sec:hardware}).

\subsection{IBM quantum hardware results}
\label{sec:hardware}

To conclude our study of the advection equation, we experimentally verify our circuits on IBM quantum hardware. Specifically, we compare results from the shallower SAA circuits with the more accurate but deeper QSP circuits in the presence of the common hardware noise. We reduce the spatial resolution for those experiments anticipating non-negligible NISQ noise. 

\begin{figure*}[!t]
    \centering
    \includegraphics[width=\columnwidth]{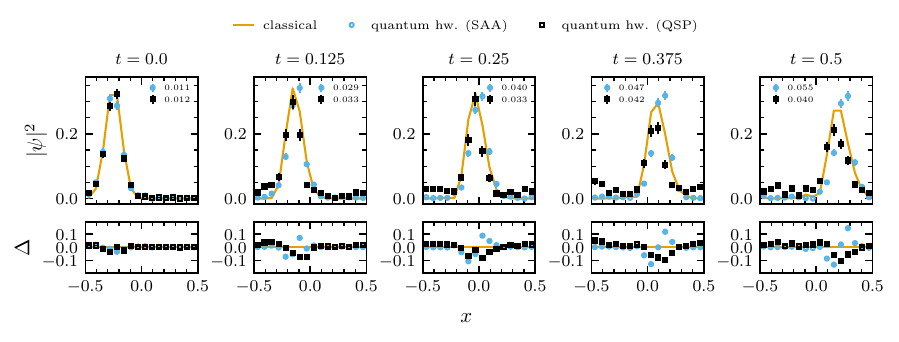}
    \caption[]{IBMQ hardware results for solving the advection equation. We solve the advection equation~\eqref{eq:advection_equation} at $t = 0, 0.125, 0.25, 0.375, 0.5$ on a classical computer (orange solid line) and on quantum hardware (blue dot for SAA~\eqref{eq:advection_solution_SAA} and black square for QSP with Jacobi-Anger expansion~\eqref{eq:advection_solution_QSP_JA}). The top panel shows $\abs{\psi}^2$ and the bottom panel shows the error compared to the classical simulation, $\Delta =  \abs{\psi}^2- \abs*{\psi^\mathrm{classical}}^2$. For the quantum results, we use IBM's \texttt{ibm\_miami} with $1,000$ shots per run. We compute and display the root mean squared error (RMSE) to evaluate the quantum circuit's overall performance. Parameters used: $n = 4$ system qubits ($n+1=5$ total qubits for QSP-based circuits), $r = 1$, and degrees $d = 0, 2, 4, 6, 8$ in \Cref{eq:advection_solution_QSP_JA} for the QSP-based circuits.
    \label{fig:advection_hardware}}
\end{figure*}

As in \Cref{sec:adv_numerics}, we use the Gaussian wave-packet from \Cref{eq:gaussian_initial} with $\mu = -0.25$ and $\sigma = 0.1$. We solve the advection equation for different times $t = 0, 0.125, 0.25, 0.375, 0.5$. Our results are summarized in \Cref{fig:advection_hardware}. The quantum hardware runs are performed on \texttt{ibm\_miami} with $n = 4$ system qubits (16 grid points) and $1,000$ shots per run. The orange solid line is for the classical solution, while the blue circle and black square are for the quantum hardware solutions with SAA and QSP, respectively. Note that we plot the wavefunction amplitude squared $\abs{\psi}^2$ in the top panel as this is the quantity we measure on hardware, and the error $\Delta = \abs{\psi}^2 - \abs*{\psi^\mathrm{classical}}^2$ in the bottom panel. For the quantum solution with QSP (total $n+1=5$ qubits), we use the Jacobi-Anger expansion in \Cref{eq:advection_solution_QSP_JA}, which reduces the circuit depth for smaller $t$'s compared to the Fourier series approach~\eqref{eq:advection_solution_qsvt}. Notably, at this system size, QSP achieves lower errors than SAA for $t\ge 0.25$ as indicated by RMSE between $\abs{\psi(x)}^2$ and $\abs{\psi^\mathrm{classical}(x)}^2$. Our RMSEs and the gate requirements are tabulated in \Cref{tab:hardware_gates}. As expected, the quantum circuits with QSP require more two-qubit gates and larger depth than the quantum circuits with SAA, but the accuracy improvement outweighs the additional noise cost due to larger circuits, resulting in smaller RMSE.

 \begin{table}[!t]
    \centering
    \caption{IBMQ hardware gate requirements and errors for solving the advection equation with $n = 4$. This table complements \Cref{fig:advection_hardware} and lists the number of two-qubit gates required, which equals the circuit depth after transpilation to IBMQ architecture for this setting. We also show the RMSE at each time $t$. Left group: SAA; right group: QSP.}
    \begin{tabular}{lcccc}
    \toprule
    & \multicolumn{2}{c}{SAA} & \multicolumn{2}{c}{QSP} \\
    \cmidrule(lr){2-3}\cmidrule(lr){4-5}
           Time $t$ & \# $2q$-gates & RMSE     & \# $2q$-gates & RMSE\\
    \midrule
    $0$ & $15$ & $0.011$ & $15$ & $0.012$ \\
    $0.125$ & $36$ & $0.029$  & $86$ & $0.033$ \\
    $0.25$  & $36$ & $0.040$ & $114$ & $0.033$ \\
    $0.375$ & $36$ & $0.047$ & $142$ & $0.042$ \\
    $0.5$ & $36$ & $0.055$ & $170$ & $0.040$ \\
    \bottomrule
    \end{tabular} \label{tab:hardware_gates}
\end{table}

 Our results verify that our quantum circuits can be used to successfully solve the advection equation even on NISQ devices. Crucially, to the best of our knowledge, we provide the first hardware implementation and experimental quantum hardware verification that QSP can be used not only to solve PDEs but also to enhance the solutions to PDEs by achieving more accurate solutions.

\section{Wave equation}\label{sec:wave}

The one-dimensional wave equation describes the propagation of disturbances through a medium in two opposite directions at a fixed speed $c$:
\begin{equation} \label{eq:wave_equation}
    \pdv[2]{\psi}{t} = c^2 \pdv[2]{\psi}{x}.
\end{equation}
Henceforth, we set $c = 1$. After discretizing space using the central finite-difference Laplacian $\Delta_x$ from \Cref{eq:laplacian}, we obtain
\begin{equation}
    \pdv[2]{\ket{\psi}}{t} = \Delta_x \ket{\psi}.
\end{equation}
In the Fourier basis, it is convenient to introduce the Hermitian operator
\begin{equation}
    S_x := \QFT \left[ 2N \sin\left(\frac{\pi \waveop}{N}\right) \right] \QFT^\dagger,
\end{equation}
which satisfies $S_x^2 = -\Delta_x$. Restricting to the nonzero-mode subspace, we define the auxiliary field
\begin{equation} \label{eq:wave_phi}
    \ket{\phi} = {\mathrm i} S_x^{-1} \pdv{\ket{\psi}}{t},
\end{equation}
and write $\ket{\Psi} \equiv (\ket{\psi}, \ket{\phi})^T$. The discrete wave equation can then be recast as
\begin{equation}\label{eq:wave_twolevel}
    \pdv{t}\begin{pmatrix} \ket{\psi} \\ \ket{\phi} \end{pmatrix}
    = -{\mathrm i}
    \begin{pmatrix}
        0 & S_x \\
        S_x & 0
    \end{pmatrix}
    \begin{pmatrix} \ket{\psi} \\ \ket{\phi} \end{pmatrix}.
\end{equation}
Equivalently, we may write a Schr\"odinger-like equation ${\mathrm i}\partial_t \ket{\Psi} = \hat{H} \ket{\Psi}$ with Hamiltonian
\begin{equation}
    \hat{H} = X \otimes S_x = (H \otimes \QFT)(Z \otimes 2N \sin{(\pi \waveop/N)})(H \otimes \QFT^\dagger),
\end{equation}
which leads to the solution
\begin{align} \label{eq:wave_equation_solution}
\begin{split}
    \ket{\Psi(x,t)} &= (H \otimes \QFT) e^{-{\mathrm i} t Z \otimes 2N \sin(\pi \waveop/N)} (H \otimes \QFT^\dagger) \ket{\Psi(x,t=0)}.
\end{split}
\end{align}
In the present manuscript we report only the small-angle approximation (SAA) circuit for the wave equation, a Laurent-polynomial QSP implementation is not included. Using $\sin(\pi \waveop/N) \approx \pi \waveop/N$, we obtain
\begin{equation} \label{eq:wave_equation_solution_SAA}
    e^{-{\mathrm i}2tNZ\otimes \sin(\pi \waveop/N)} \approx e^{-{\mathrm i}2\pi t Z \otimes \waveop} = e^{{\mathrm i}\pi t Z_0} \prod\limits^n_{\beta =1} e^{{\mathrm i}\pi t 2^{n-\beta} Z_0 \otimes Z_\beta}.
\end{equation}
The right-hand side is a product of a one-qubit Pauli-Z rotation gate and $n$ two-qubit ZZ rotations, which yields a depth-$n$ quantum circuit. Combining \Cref{eq:wave_equation_solution_SAA} with \Cref{eq:wave_equation_solution} gives the quantum circuit used to solve the discretized wave equation. This solution requires preparing the initial state $\ket{\Psi(x,t=0)}$ from $\psi(x,t=0)$ and $\partial_t\psi(x,t=0)$.

\subsection{Numerical results}

\begin{figure}[htbp]
    \centering
    \includegraphics[width=\columnwidth]{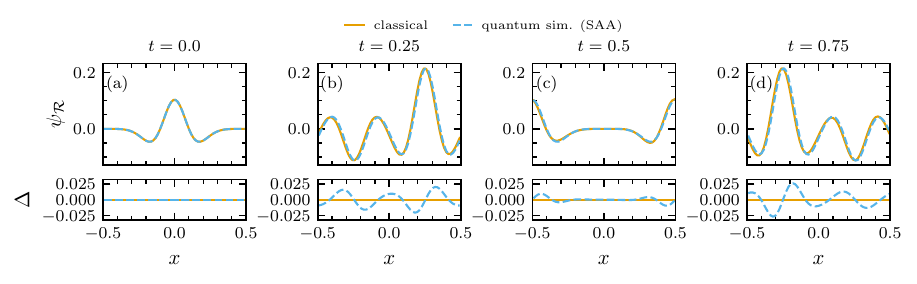}
    \caption[]{Numerical solutions to the wave equation. Here, we initialize $\psi(t=0)$ with a Ricker wavelet~\eqref{eq:ricker_wavelet} and evaluate at different times $t$. The real part of the solution $\psi_\mathcal{R}$ is plotted on the top panel, while the bottom panel shows the difference from the classical solution. Parameters used: $n = 6$ qubits and $c = 1$.  \label{fig:wave_simulation_2x2}}
\end{figure} 

For the wave equation~\eqref{eq:wave_equation}, we use the SAA circuit from \Cref{eq:wave_equation_solution_SAA} to solve the discretized problem. For our initial conditions $\psi(x,t=0)$ and $\partial_t \psi(x,t=0)$, we use the Ricker wavelet as done in Ref.~\cite{wright2024noisy}:
\begin{equation} \label{eq:ricker_wavelet}
    \psi(x,t=0) = \frac{2}{\sqrt{3 \sigma} \pi^{1/4}} \left[1- \left(\frac{x-\mu}{\sigma} \right)^2\right] e^{-\frac{(x-\mu)^2}{2\sigma^2}},
\end{equation}
where we set $\partial_t \psi(x, t=0) = \partial_x \psi(x,t=0)$, $\mu = 0$, and $\sigma = 0.1$.  
We verify the SAA-based solution in \Cref{fig:wave_simulation_2x2} using statevector simulations and compare to classical solver for the same discretized wave equation and initial data. A key difference from Ref.~\cite{wright2024noisy} is that our initial condition for $\partial_t\psi(x,t=0)$ is nontrivial, whereas the authors of that reference set $\partial_t\psi(x,t=0) = 0$. Specifically, implementing $\partial_t \psi(x,t=0)$ means simply setting $\ket{\phi} = \ket{0}$. On the other hand, a non-zero $\partial_t \psi(x,t=0)$ requires implementing $\ket{\phi}$ using \Cref{eq:wave_phi}.

\section{Poisson equation}\label{sec:poisson}
The one-dimensional Poisson equation describes the potential field generated by a source distribution and is given by
\begin{equation} \label{eq:poisson_equation}
    \pdv[2]{\psi}{x} = \rho,
\end{equation}
where $\rho=\rho(x)$ is the (unnormalized) source term. For periodic boundary conditions, the discretized problem on the $N$-point grid is
\begin{equation} \label{eq:poisson_discrete_system}
    \Delta_x {\psi} = \ket{\rho},
\end{equation}
where we assume the source term $\ket{\rho}$ to be normalized and denote the (unnormalized) solution ${\psi}\in\mathbb{C}^N$ on the grid $x_j$.
We explicitly do not write an identity between normalized quantum states. We will derive a Laurent-polynomial QSP construction that prepares $\ket{\psi} \propto {\psi}$ up to the additional normalization factor $\norm{\rho}$ and any additional subnormalization introduced by the QSP construction. A (unitary) SAA implementation is not included as the Poisson equation does not allow for such a representation.

Since the periodic discrete Laplacian $\Delta_x$ from \Cref{eq:laplacian} has the zero mode in its kernel, its inverse does not exist. We therefore work with the Moore-Penrose pseudoinverse $\Delta_x^+$. Let
\begin{equation}
    \Pi_0 := \dyad*{\tilde{k}=0}
\end{equation}
denote the projector onto the zero Fourier mode. Then
\begin{equation}
    \Delta_x \Delta_x^+ = \Delta_x^+ \Delta_x = I - \Pi_0.
\end{equation}
Thus, for an arbitrary periodic source $\ket{\rho}$, the pseudoinverse solution, ${\psi} = \Delta_x^+ \ket{\rho}$,
satisfies the projected equation
\begin{equation} \label{eq:poisson_projected}
    \Delta_x {\psi} = (I-\Pi_0)\ket{\rho}.
\end{equation}
If $\ket{\rho}$ has zero discrete mean, equivalently $\Pi_0 \ket{\rho}=0$, then \Cref{eq:poisson_projected} reduces to the usual periodic Poisson equation $\Delta_x {\psi} = \ket{\rho}$. We therefore work on the mean-zero subspace and fix the solution by the gauge condition $\braket*{\tilde{k}=0}{\psi}=0$.
The resulting unnormalized solution vector is
\begin{equation} \label{eq:poisson_solution}
    {\psi} = \Delta_x^+ \ket{\rho}
    = \QFT \cdot \Lambda^+ \cdot \QFT^\dagger \ket{\rho},
\end{equation}
and
\begin{equation} \label{eq:poisson_pseudoinverse}
    \Lambda^+ \ket{k} =
    \begin{cases}
        \displaystyle \left[-4N^2 \sin^2\left(\frac{\pi \tilde{k}}{N}\right)\right]^{-1} \ket{k},
        & \tilde{k} \neq 0, \\[1ex]
        0, & \tilde{k} = 0.
    \end{cases}
\end{equation}
It follows that the Fourier-space pseudoinverse $\Lambda^+$ admits the Laurent expansion
\begin{equation}
    \Lambda^+ = \sum\limits^{N/2-1}_{m=-N/2} c_m e^{{\mathrm i}2\pi m\waveop/N},
\end{equation}
with coefficients
\begin{equation} \label{eq:poisson_QSP_coefficients}
    c_m = \frac{1}{N} \sum\limits^{N/2-1}_{\substack{\tilde{k}=-N/2 \\ \tilde{k}\neq 0}} \left[-4N^2 \sin^2\left(\frac{\pi \tilde{k}}{N}\right)\right]^{-1} e^{-{\mathrm i}2\pi m \tilde{k}/N}.
\end{equation}
In the language of \Cref{subsec:QSP}, we implement this Laurent polynomial using QSP with $U = e^{{\mathrm i}2\pi \waveop/N}$ to encode the Fourier-space pseudoinverse of the Laplacian $\Lambda^+$. The $\ket{0}$-post-selected quantum output is a normalized state $\ket{\psi}  \propto {\psi}$, and the overall scale of the unnormalized solution vector is recovered by multiplying $\norm{\rho}$ with the QSP scaling factor described after \Cref{eq:QSP_on_initial_state}.

\subsection{Dirichlet boundary conditions via odd extension}

The periodic pseudoinverse solver above can be used to impose Dirichlet boundary conditions, which sets fixed values to the solution at the boundaries (see \Cref{eq:poisson_dirichlet_half}), by reducing a problem on a half-interval $(-\tfrac{1}{2}, 0)$ to a mean-zero periodic problem on a doubled grid $(-\tfrac{1}{2}, \tfrac{1}{2})$. Consider
\begin{equation} \label{eq:poisson_dirichlet_half}
    \pdv[2]{\tilde{\psi}}{x} = \tilde{\rho}(x),
    \qquad x \in \Omega_- := \left(-\tfrac{1}{2},0\right),
    \qquad
    \tilde{\psi}\left(-\tfrac{1}{2}\right)=0,
    \quad
    \tilde{\psi}(0)=\beta.
\end{equation}
In contrast to the periodic problem, no mean-zero condition is imposed on the original source $\tilde{\rho}$ on $\Omega_-$. Then, we solve for $\psi$ in the doubled grid $(-\tfrac{1}{2}, \tfrac{1}{2})$, where the true solution can be recovered using $\tilde{\psi} = R_- \psi$, which lies in $\Omega_-$. $R_-:\mathbb{C}^{N}\to\mathbb{C}^{\tilde N}$ restricts to the first $\tilde N$ grid points.

We first consider the homogeneous case $\beta=0$. Define the odd extension of $\tilde{\rho}$ to the doubled domain $\Omega:=(-\tfrac{1}{2},\tfrac{1}{2})$ by
\begin{equation} \label{eq:poisson_odd_extension_cont}
    \rho(x) =
    \begin{cases}
        \tilde{\rho}(x), & x \in \left(-\frac{1}{2},0\right), \\[0.5ex]
        -\tilde{\rho}(-x), & x \in \left(0,\frac{1}{2}\right).
    \end{cases}
\end{equation}
Since $\rho$ is odd, it has zero mean on $\Omega$, so the mean-zero periodic auxiliary problem
\begin{equation} \label{eq:poisson_periodic_aux}
    \pdv[2]{{\psi}}{x}=\rho(x),
    \qquad x \in \Omega,
    \qquad
    \int_{-1/2}^{1/2} \psi(x)\, \dd x = 0
\end{equation}
is well posed. Its solution is also odd, hence $\psi(-\tfrac{1}{2})=\psi(0)=0$, and restricting $\psi$ to $\Omega_-$ yields the homogeneous Dirichlet solution.

\begin{figure}[!t]
    \centering
    \includegraphics[width=\columnwidth]{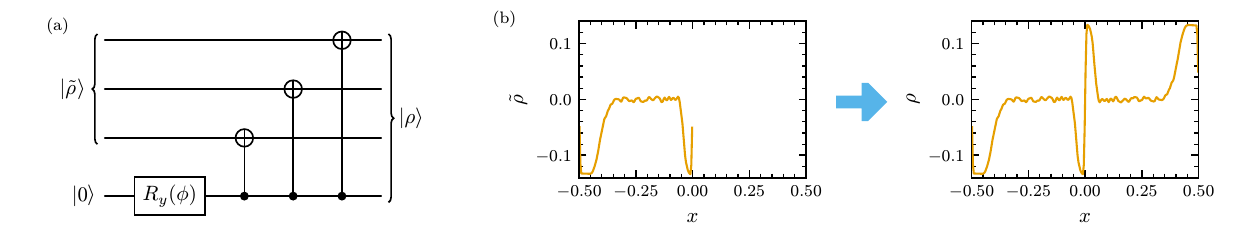}
    \caption[]{Circuit construction for the odd extension isometry $V_{\text{odd}}$ used to reduce a Dirichlet Poisson problem on the half-interval $\Omega_- = (-\tfrac{1}{2},0)$ to a periodic auxiliary problem on the doubled interval $\Omega = (-\tfrac{1}{2},\tfrac{1}{2})$. Given discretized data $\ket{\tilde{\rho}}$ on $\Omega_-$, panel (a) implements the discrete odd-extension isometry $V_{\mathrm{odd}}$ from \Cref{eq:odd_isometry}, producing the extended source $\ket{\rho}$. Panel (b) illustrates the resulting odd periodic source on the doubled domain. \label{fig:poisson_supp}}
\end{figure}

Next, we discuss how the odd source term $\ket{\rho}$ can be implemented starting from the source term $\ket{\tilde{\rho}}$ on the half-interval. Let $\tilde{N}=2^{\tilde n}$ (where $\Omega_{-}$ is represented by $\tilde{N} = 2^{\tilde n}$ grid points) and $N=2\tilde N=2^n$, $n=\tilde n + 1$. We write $\tilde{x}_j=x_j$ for $j\in[\tilde N]$, i.e., the first $\tilde N$ points of the doubled periodic grid from \Cref{eq:xgrid}. Writing $\bar{j}:=\tilde N-1-j$,
the odd extension can be implemented by the isometry
\begin{equation}
    V_{\mathrm{odd}}\ket{j}
    =
    \frac{1}{\sqrt2}\left(\ket{0}\ket{j}-\ket{1}\ket{\bar j}\right),
    \qquad
    \ket{\rho}=V_{\mathrm{odd}}\ket{\tilde\rho}.
\label{eq:odd_isometry}
\end{equation}
This isometry is implemented by adding one ancilla qubit in the state $\ket{-}$ and applying Pauli-$X$ to each system qubit encoding $\ket{\tilde{\rho}}$ controlled on the ancilla being in the state $\ket{1}$. Indeed, bitwise complementation of the lower register maps $j$ to $\bar j=\tilde N-1-j$. This construction is summarized in~\Cref{fig:poisson_supp}.

Afterwards, the solution $\ket{\psi} \propto \psi$ on the double-interval $\Omega$ can be prepared using the QSP implementation of the pseudoinverse $\Delta_x^+$ as outlined above. The physical solution on the half-interval $\Omega_{-}$ for the source $\ket{\tilde{\rho}}$ is recovered as
\begin{equation}
    \tilde{{\psi}}
    =
    \norm{{\rho}} \,R_-\ket{\psi}.
\end{equation}

To impose the non-homogeneous boundary value $\tilde{\psi}(0)=\beta$, we add the boundary-lifting function
\begin{equation} \label{eq:poisson_boundary_lift_cont}
    \ell_\beta(x)=2\beta\left(x+\frac{1}{2}\right),
    \qquad x\in\Omega_-,
\end{equation}
which satisfies $\ell_\beta''(x)=0$, $\ell_\beta(-\tfrac{1}{2})=0$, and $\ell_\beta(0)=\beta$. Its discrete counterpart is
\begin{equation} \label{eq:poisson_boundary_lift_disc}
    \tilde{\ell}_\beta
    =
    \sum_{j=0}^{\tilde N-1}
    2\beta\left(\tilde{x}_j+\frac{1}{2}\right)\ket{j}.
\end{equation}
Hence the final discretized Dirichlet solution for non-homogeneous boundary conditions can be retrieved as
\begin{equation} \label{eq:poisson_dirichlet_final}
    \tilde{\psi} = s_{\rm QSP}\norm{\rho}\,R_-\Delta_x^+V_{\mathrm{odd}}\ket{\tilde\rho} + \tilde{{\ell}}_\beta,
\end{equation}
where $s_{\rm QSP}$ is the QSP success probability factor, $\Delta_x^+ V_{\text{odd}}\ket{\tilde\rho}$ are evaluated on the quantum computer, and the remaining operations can be computed efficiently in classical post-processing.

\subsection{Numerical results for periodic boundary conditions}

\begin{figure}[!t]
    \centering
    \includegraphics[width=\columnwidth]{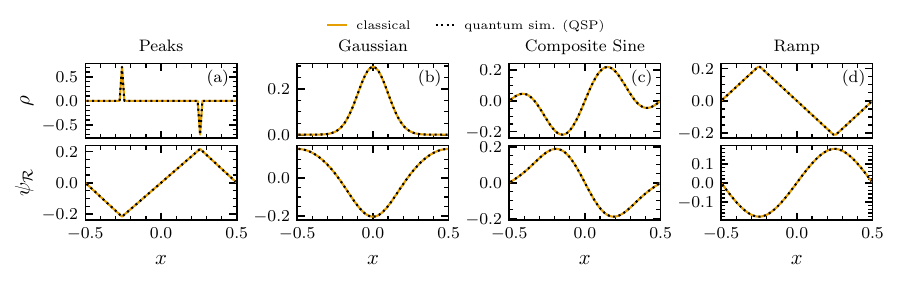}
    \caption[]{Numerical solutions to the Poisson equation (\Cref{eq:poisson_equation}). We solve for varying source terms: (a) peaks, (b) Gaussian, (c) composite sine, and (d) ramp from \Cref{eq:poisson_sources}. Classical solutions (quantum circuit solutions using QSP) are depicted as orange solid lines (black dotted lines), denoted as $\psi_\mathcal{R}$. Parameters used: $n = 6$ qubits for quantum circuit solutions.
    \label{fig:poisson_simulation_2x2}}
\end{figure}

We first numerically verify our quantum circuit solution (\Cref{eq:poisson_solution}) to the Poisson equation (\Cref{eq:poisson_equation}) for periodic boundary conditions. We consider the following source terms $\rho(x)$ that satisfy periodic boundary conditions:
\begin{align} \label{eq:poisson_sources}
\begin{split}
& \text{Peaks: } \,  \rho(x) = \begin{cases}
    1 \quad \text{if } \, x = x_{N/4-1}, \\
    -1 \quad \text{if } \, x = x_{3N/4}, \\
    0 \quad \text{otherwise}
\end{cases} \\
& \text{Gaussian: } \rho(x)= \frac{1}{\sqrt{2 \pi}} e^{-x^2/2}, \\
& \text{Composite Sine: } \rho(x) = \sin(2\pi x) + \sin(4\pi x), \\
& \text{Ramp: } \rho(x) = \begin{cases}
     x+\frac{1}{2} \quad \text{if } \, -\frac{1}{2}\leq x < -\frac{1}{4}, \\
     -x \quad \text{if } \, -\frac{1}{4} \leq x \leq \frac{1}{4}, \\
     x - \frac{1}{2} \quad \text{if } \, \frac{1}{4} < x \leq \frac{1}{2}, \\
     0 \quad \text{otherwise}.
\end{cases}
\end{split}
\end{align}
We solve the projected equation (\Cref{eq:poisson_projected}), where we neglect any non-zero mean of the source term $\rho$ (e.g., the Gaussian wavepacket has non-zero mean). These solutions $\psi_\mathcal{R}(x)$ are plotted in \Cref{fig:poisson_simulation_2x2} alongside classical PDE solver results and quantum circuit solutions using QSP. These results confirm our numerical implementations for solving the Poisson equation.

\begin{figure}[!t]
    \centering
    \includegraphics[width=0.6\columnwidth]{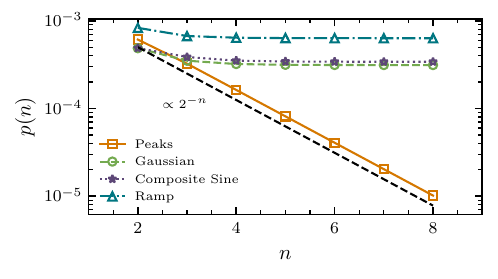}
    \caption[]{Probability of successful measurements for solving the Poisson equation. We compute the probability that the ancilla qubit measures $0$ for the Poisson solution circuit with $n$ qubits, denoted as $p(n)$, using statevector simulations of the quantum circuits. Different markers (and colors) refer to different source terms in \Cref{eq:poisson_sources}. The dashed black line indicates $2^{-n}$ scaling. \label{fig:poisson_probability}}
\end{figure} 

A downside of the QSP-based method is post-selection: only shots where the ancilla qubit measures $0$ contribute to the solution. With \Cref{eq:poisson_solution}, the success probability for solving the Poisson equation scales as $\Omega(N^{-4}) = \Omega(2^{-4n})$ in the worst case, which arises from the fact that the smallest value (in absolute value) of the pseudoinverse $\Lambda^+$~\eqref{eq:poisson_pseudoinverse} is $-N^2/4$ from $\tilde{k} = -N/2$ and thus $p(n) = \Omega(N^{-4})$. For example, this sort of worst-case scaling can arise if $\ket{\rho}$ is a plane wave of this wavenumber $\tilde{k} = -N/2$. These probabilities are $p(n) = \alpha^2$ (see \Cref{eq:QSP_on_initial_state}), where $n$ is the number of qubits, and depend on the source term, as shown in \Cref{fig:poisson_probability}. Interestingly, $p(n)$ saturates for sufficiently large $n \gtrsim 5$ for all source terms other than ``peaks''. As explained in Ref.~\cite{lubasch2025}, smooth initial functions $\rho$ lead to constant success probability with $\Omega(\abs*{\tilde{k}_\mathrm{max}}^2)$, where $\abs*{\tilde{k}_\mathrm{max}}$ is the maximum wavenumber such that $\abs*{\braket*{\tilde{k}}{\rho}}^2$ has non-vanishing overlap. As this maximum wavenumber is independent of $n$, $p(n)$ saturates this theoretical value. For ``peaks'', however, $p(n)$ $p(n)$ continues to decrease as $n$ increases, with $p(n) \propto 2^{-n}$. This is due to the fact that the QFT of $\ket{\rho}$ leads to $\ket*{\tilde{k}}$ that represents an equal superposition of all wavenumbers, which leads to $\Omega(N^{-1})$ scaling for $p(n)$ as in \Cref{fig:poisson_probability}:
\begin{equation}
    p(n) =\sum_m \abs{c_m}^2 \approx N \left[ \frac{1}{N^2} \frac{1}{16N^4} \sum_{\tilde{k}} \csc^4{\left(\frac{\pi \tilde{k}}{N}\right)} \right] = \Omega\left(\frac{N^4}{N^5}\right) = \Omega(N^{-1}),
\end{equation}
where we use $c_m$ defined in \Cref{eq:poisson_QSP_coefficients}.

Nevertheless, the probability of successful measurements is on the order of $10^{-4}$, significantly prohibiting real hardware implementations with NISQ devices. Assuming independent random measurements, the signal-to-noise ratio $\sigma(p)/p$, where $\sigma(p) = \sqrt{p(1-p)/M}$ and $M$ is the number of measurements. Achieving a signal-to-noise ratio of $10\%$ requires more than $10^6$ measurements for $p = 10^{-4}$. This does not even account for additional measurements needed to determine individual spatial points of $\psi(x)$, which further exacerbates the situation.

\subsection{Numerical example with non-homogeneous Dirichlet data}
We verify our quantum circuit approach for a Poisson solve with non-ho\-mo\-ge\-neous boundary conditions, where the source terms are obtained from high-fidelity physics simulations. In particular, we perform high-fidelity Particle-In-Cell/Monte-Carlo Collisions (PIC/MCC)~\cite{birdsall} simulations of a capacitively coupled plasma (CCP), a prototypical low-temperature discharge widely used in semiconductor chip fabrication. 
In the electrostatic PIC/MCC method, the plasma is represented by computational particles (each particle represents a large number of electrons/ions) that carry charge. The charge is deposited onto a spatial grid using a cloud-in-cell approach to obtain charge density, which is used as a source term for the Poisson equation to obtain the electrostatic potential with appropriate boundary conditions. The resulting electric field is used to accelerate particles and advance their positions and velocities. The particle velocities are also affected by MCC collisions (including ionization, elastic scattering with constant background gas). We used the open-source WarpX code~\cite{myers2021} to perform PIC/MCC simulations of a 1D CCP set-up consistent with the well-established benchmark~\cite{turner2013}. In this set-up, the 1D domain discretized with 128 grid cells is grounded on one side and the other boundary is driven by a sinusoidal radio-frequency voltage with 450~V amplitude and frequency $f= $ 13.56~MHz, i.e., the boundary conditions are $(0\ V(t))$. The time-dependent boundary condition at the powered electrode is therefore given by $450\sin(2\pi f t)$~V, where $t$ is the physical time, and the simulation employed 400 timesteps per RF cycle. The resulting self-consistent charge density profiles are used as source terms for the Poisson equation solved using our quantum circuit approach.

\begin{figure}[!t]
    \centering
    \includegraphics[width=1\columnwidth]{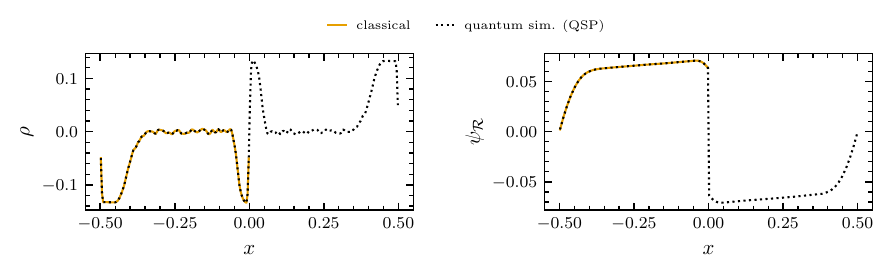}
    \caption[]{Numerical example for the Poisson equation with non-homogeneous Dirichlet boundary conditions on the half-interval $\Omega_- = (-0.5,0)$. The boundary data satisfy $\tilde{\psi}(-0.5)=0$ and $\tilde{\psi}(0)=450 \sin(2\pi f t)$~V, where $f = 13.56$~MHz and $t = 18.4$~ns ($100$-th timestep). Parameters used: $n = 8$ qubits for the QSP circuit solution. \label{fig:poisson_real_data}}
\end{figure}

In \Cref{fig:poisson_real_data}, we apply \Cref{eq:poisson_dirichlet_final} to the source term $\tilde{\rho}$ in the half-interval $\Omega_-$, obtained using the charge density values derived from the PIC/MCC simulations at $t = 18.4$~ns. The input data have $\tilde{N}=128$ grid points, so the auxiliary periodic problem is posed on $N=256$ grid points and thus uses $n=8$ qubits with one additional ancilla qubit for QSP. We first construct the odd extension $V_{\mathrm{odd}}\ket{\tilde{{\rho}}}$, solve the periodic auxiliary problem on the doubled grid, restrict the result back to the physical half-interval, and finally add the boundary-lifting vector $\tilde{{\ell}}_\beta$. The resulting statevector solution agrees with the classically computed discretized solution on the physical domain, providing a proof-of-concept implementation of the non-homogeneous Dirichlet-boundary construction.

\section{Conclusion}

In this paper, we develop and demonstrate an end-to-end implementation (see \texttt{qsp4pde}~\cite{qsp4pde}) of quantum circuits designed to solve linear partial differential equations using the quantum Fourier transform and quantum signal processing, building on Ref.~\cite{lubasch2025}. Our three-fold contributions include: first, the development of explicit compiled quantum circuits that we successfully benchmarked through both numerical statevector simulations and experimental runs on IBM quantum hardware; second, the experimental demonstration that QSP-based quantum circuits can achieve more accurate solutions for the advection equation than small-angle approximation methods for relatively small systems but under realistic NISQ hardware constraints; and third, the generalization of our quantum simulation methodology to accommodate non-homogeneous Dirichlet boundary conditions, which we validated by numerically solving the Poisson equation using self-consistent charge density profiles derived from high-fidelity plasma physics simulations. Our results showcase that advanced quantum signal processing techniques can enhance the accuracy and scope of PDE solvers on near-term quantum devices. Nonetheless, several limitations remain. The algorithms use amplitude-encoded input data, so state preparation and readout costs must be included when evaluating particular applications. In addition, the Poisson solver relies on post-selection, and the success probability depends strongly on the spectral content of the source term. Reducing QSP polynomial degrees, mitigating post-selection overhead, and extending the approach to higher-dimensional or less structured discretizations are important directions for future work. Overall, the results provide an implementation-level benchmark for studying how QSP-based PDE solvers behave after compilation, simulation, and execution on present-day quantum hardware.

\section*{Acknowledgments}
This research used IBM quantum hardware, accessed using resources of the National Energy Research Scientific Computing Center (NERSC), a Department of Energy User Facility. This research used the open-source \href{https://blast-warpx.github.io}{particle-in-cell code WarpX} for the plasma simulations. Primary WarpX contributors are with LBNL, LLNL, CEA-LIDYL, SLAC, DESY, CERN, Helion Energy, and TAE Technologies.
We acknowledge all WarpX contributors.

\bibliographystyle{siamunsrt}
\bibliography{refs}

@article{lubasch2025,
  title = {Quantum circuits for partial differential equations in Fourier space},
  author = {Lubasch, Michael and Kikuchi, Yuta and Wright, Lewis and Mc Keever, Conor},
  journal = {Phys. Rev. Res.},
  volume = {7},
  issue = {4},
  pages = {043326},
  numpages = {20},
  year = {2025},
  month = {Dec},
  publisher = {American Physical Society},
  doi = {10.1103/tbzc-w9x8},
  url = {https://link.aps.org/doi/10.1103/tbzc-w9x8}
}

@article{motlagh2024generalized,
  title = {Generalized Quantum Signal Processing},
  author = {Motlagh, Danial and Wiebe, Nathan},
  journal = {PRX Quantum},
  volume = {5},
  issue = {2},
  pages = {020368},
  numpages = {16},
  year = {2024},
  month = {Jun},
  publisher = {American Physical Society},
  doi = {10.1103/PRXQuantum.5.020368},
  url = {https://link.aps.org/doi/10.1103/PRXQuantum.5.020368}
}

@misc{yamamotoRobustAngleFinding2024,
  title = {Robust {{Angle Finding}} for {{Generalized Quantum Signal Processing}}},
  author = {Yamamoto, Shuntaro and Yoshioka, Nobuyuki},
  year = 2024,
  month = sep,
  number = {arXiv:2402.03016},
  eprint = {2402.03016},
  primaryclass = {quant-ph},
  publisher = {arXiv},
  doi = {10.48550/arXiv.2402.03016},
}

@article{martynGrandUnificationQuantum2021,
  title = {Grand {{Unification}} of {{Quantum Algorithms}}},
  author = {Martyn, John M. and Rossi, Zane M. and Tan, Andrew K. and Chuang, Isaac L.},
  year = 2021,
  month = dec,
  journal = {PRX Quantum},
  volume = {2},
  number = {4},
  pages = {040203},
  publisher = {American Physical Society},
  doi = {10.1103/PRXQuantum.2.040203},
}

@inproceedings{gilyen,
author = {Gily\'{e}n, Andr\'{a}s and Su, Yuan and Low, Guang Hao and Wiebe, Nathan},
title = {Quantum singular value transformation and beyond: exponential improvements for quantum matrix arithmetics},
year = {2019},
isbn = {9781450367059},
publisher = {Association for Computing Machinery},
address = {New York, NY, USA},
url = {https://doi.org/10.1145/3313276.3316366},
doi = {10.1145/3313276.3316366},
booktitle = {Proceedings of the 51st Annual ACM SIGACT Symposium on Theory of Computing},
pages = {193–204},
numpages = {12},
location = {Phoenix, AZ, USA},
series = {STOC 2019}
}

@article{wright2024noisy,
  title = {Noisy intermediate-scale quantum simulation of the one-dimensional wave equation},
  author = {Wright, Lewis and Mc Keever, Conor and First, Jeremy T. and Johnston, Rory and Tillay, Jeremy and Chaney, Skylar and Rosenkranz, Matthias and Lubasch, Michael},
  journal = {Phys. Rev. Res.},
  volume = {6},
  issue = {4},
  pages = {043169},
  numpages = {10},
  year = {2024},
  month = {Nov},
  publisher = {American Physical Society},
  doi = {10.1103/PhysRevResearch.6.043169},
  url = {https://link.aps.org/doi/10.1103/PhysRevResearch.6.043169}
}

@misc{coppersmith2002approximatefouriertransformuseful,
      title={An approximate Fourier transform useful in quantum factoring}, 
      author={D. Coppersmith},
      year={2002},
      eprint={quant-ph/0201067},
      archivePrefix={arXiv},
      primaryClass={quant-ph},
      url={https://arxiv.org/abs/quant-ph/0201067}, 
}

@book{Nielsen2010,
  author    = {Nielsen, Michael A. and Chuang, Isaac L.},
  title     = {Quantum Computation and Quantum Information: 10th Anniversary Edition},
  publisher = {Cambridge University Press},
  year      = {2010},
  edition   = {10th Anniversary},
  isbn      = {978-1107002173}
}

@article{harrow09,
  title = {Quantum Algorithm for Linear Systems of Equations},
  author = {Harrow, Aram W. and Hassidim, Avinatan and Lloyd, Seth},
  journal = {Phys. Rev. Lett.},
  volume = {103},
  issue = {15},
  pages = {150502},
  numpages = {4},
  year = {2009},
  month = {10},
  publisher = {American Physical Society},
  doi = {10.1103/PhysRevLett.103.150502},
  url = {https://link.aps.org/doi/10.1103/PhysRevLett.103.150502}
}

@article{Berry_2014,
    doi = {10.1088/1751-8113/47/10/105301},
    url = {https://doi.org/10.1088/1751-8113/47/10/105301},
    year = {2014},
    month = {feb},
    publisher = {IOP Publishing},
    volume = {47},
    number = {10},
    pages = {105301},
    author = {Berry, Dominic W},
    title = {High-order quantum algorithm for solving linear differential equations},
    journal = {Journal of Physics A: Mathematical and Theoretical},
}

@article{Childs2020,
  title = {Quantum Spectral Methods for Differential Equations},
  volume = {375},
  ISSN = {1432-0916},
  url = {http://dx.doi.org/10.1007/s00220-020-03699-z},
  DOI = {10.1007/s00220-020-03699-z},
  number = {2},
  journal = {Communications in Mathematical Physics},
  publisher = {Springer Science and Business Media LLC},
  author = {Childs,  Andrew M. and Liu,  Jin-Peng},
  year = {2020},
  month = feb,
  pages = {1427–1457}
}

@article{Childs2021,
  doi = {10.22331/q-2021-11-10-574},
  url = {https://doi.org/10.22331/q-2021-11-10-574},
  title = {High-precision quantum algorithms for partial differential equations},
  author = {Childs, Andrew M. and Liu, Jin-Peng and Ostrander, Aaron},
  journal = {{Quantum}},
  issn = {2521-327X},
  publisher = {{Verein zur F{\"{o}}rderung des Open Access Publizierens in den Quantenwissenschaften}},
  volume = {5},
  pages = {574},
  month = nov,
  year = {2021}
}

@misc{lloyd2020,
      title={Quantum algorithm for nonlinear differential equations}, 
      author={Seth Lloyd and Giacomo De Palma and Can Gokler and Bobak Kiani and Zi-Wen Liu and Milad Marvian and Felix Tennie and Tim Palmer},
      year={2020},
      eprint={2011.06571},
      archivePrefix={arXiv},
      primaryClass={quant-ph},
      url={https://arxiv.org/abs/2011.06571}, 
}

@article{liu21,
    author = {Jin-Peng Liu  and Herman Øie Kolden  and Hari K. Krovi  and Nuno F. Loureiro  and Konstantina Trivisa  and Andrew M. Childs },
    title = {Efficient quantum algorithm for dissipative nonlinear differential equations},
    journal = {Proceedings of the National Academy of Sciences},
    volume = {118},
    number = {35},
    pages = {e2026805118},
    year = {2021},
    doi = {10.1073/pnas.2026805118},
    URL = {https://www.pnas.org/doi/abs/10.1073/pnas.2026805118},
}

@misc{tennie2024,
      title={Solving nonlinear differential equations on Quantum Computers: A Fokker-Planck approach}, 
      author={Felix Tennie and Luca Magri},
      year={2024},
      eprint={2401.13500},
      archivePrefix={arXiv},
      primaryClass={quant-ph},
      url={https://arxiv.org/abs/2401.13500}, 
}

@article{Shi24,
  title = {Quantum Simulation of Partial Differential Equations via Schr\"odingerization},
  author = {Jin, Shi and Liu, Nana and Yu, Yue},
  journal = {Phys. Rev. Lett.},
  volume = {133},
  issue = {23},
  pages = {230602},
  numpages = {6},
  year = {2024},
  month = {12},
  publisher = {American Physical Society},
  doi = {10.1103/PhysRevLett.133.230602},
  url = {https://link.aps.org/doi/10.1103/PhysRevLett.133.230602}
}

@article{hu2024quantum,
  title={Quantum circuits for partial differential equations via Schr{\"o}dingerisation},
  author={Hu, Junpeng and Jin, Shi and Liu, Nana and Zhang, Lei},
  journal={Quantum},
  volume={8},
  pages={1563},
  year={2024},
  publisher={Verein zur F{\"o}rderung des Open Access Publizierens in den Quantenwissenschaften},
  doi={10.22331/q-2024-12-12-1563},
}

@article{jin2025quantum,
    title = {Quantum circuits for the heat equation with physical boundary conditions via Schrödingerization},
    journal = {Journal of Computational Physics},
    volume = {538},
    pages = {114138},
    year = {2025},
    issn = {0021-9991},
    doi = {https://doi.org/10.1016/j.jcp.2025.114138},
    url = {https://www.sciencedirect.com/science/article/pii/S0021999125004218},
    author = {Shi Jin and Nana Liu and Yue Yu},
}

@article{lubasch19,
  title = {Variational quantum algorithms for nonlinear problems},
  author = {Lubasch, Michael and Joo, Jaewoo and Moinier, Pierre and Kiffner, Martin and Jaksch, Dieter},
  journal = {Phys. Rev. A},
  volume = {101},
  issue = {1},
  pages = {010301},
  numpages = {7},
  year = {2020},
  month = {1},
  publisher = {American Physical Society},
  doi = {10.1103/PhysRevA.101.010301},
  url = {https://link.aps.org/doi/10.1103/PhysRevA.101.010301}
}

@article{pool24,
  title = {Nonlinear dynamics as a ground-state solution on quantum computers},
  author = {Pool, Albert J. and Somoza, Alejandro D. and Mc Keever, Conor and Lubasch, Michael and Horstmann, Birger},
  journal = {Phys. Rev. Res.},
  volume = {6},
  issue = {3},
  pages = {033257},
  numpages = {19},
  year = {2024},
  month = {9},
  publisher = {American Physical Society},
  doi = {10.1103/PhysRevResearch.6.033257},
  url = {https://link.aps.org/doi/10.1103/PhysRevResearch.6.033257}
}

@article{kryiienko2021,
  title = {Solving nonlinear differential equations with differentiable quantum circuits},
  author = {Kyriienko, Oleksandr and Paine, Annie E. and Elfving, Vincent E.},
  journal = {Phys. Rev. A},
  volume = {103},
  issue = {5},
  pages = {052416},
  numpages = {22},
  year = {2021},
  month = {5},
  publisher = {American Physical Society},
  doi = {10.1103/PhysRevA.103.052416},
  url = {https://link.aps.org/doi/10.1103/PhysRevA.103.052416}
}

@article{jaksch2023variational,
    author = {Jaksch, Dieter and Givi, Peyman and Daley, Andrew J. and Rung, Thomas},
    title = {Variational Quantum Algorithms for Computational Fluid Dynamics},
    journal = {AIAA Journal},
    volume = {61},
    number = {5},
    pages = {1885-1894},
    year = {2023},
    doi = {10.2514/1.J062426},
    URL = {https://doi.org/10.2514/1.J062426},
}

@misc{camps2025quantumcomputingtechnologyroadmaps,
      title={Quantum Computing Technology Roadmaps and Capability Assessment for Scientific Computing -- An analysis of use cases from the NERSC workload}, 
      author={Daan Camps and Ermal Rrapaj and Katherine Klymko and Hyeongjin Kim and Kevin Gott and Siva Darbha and Jan Balewski and Brian Austin and Nicholas J. Wright},
      year={2025},
      eprint={2509.09882},
      archivePrefix={arXiv},
      primaryClass={quant-ph},
      url={https://arxiv.org/abs/2509.09882}, 
}

@article{wang2021noise,
  title={Noise-induced barren plateaus in variational quantum algorithms},
  author={Wang, Samson and Fontana, Enrico and Cerezo, Marco and Sharma, Kunal and Sone, Akira and Cincio, Lukasz and Coles, Patrick J},
  journal={Nature communications},
  volume={12},
  number={1},
  pages={6961},
  year={2021},
  publisher={Nature Publishing Group UK London},
  doi={10.1038/s41467-021-27045-6},
  url={https://doi.org/10.1038/s41467-021-27045-6},
}

@article{larocca2025barren,
  title={Barren plateaus in variational quantum computing},
  author={Larocca, Martin and Thanasilp, Supanut and Wang, Samson and Sharma, Kunal and Biamonte, Jacob and Coles, Patrick J and Cincio, Lukasz and McClean, Jarrod R and Holmes, Zo{\"e} and Cerezo, Marco},
  journal={Nature Reviews Physics},
  volume={7},
  number={4},
  pages={174--189},
  year={2025},
  publisher={Nature Publishing Group UK London},
  doi={10.1038/s42254-025-00813-9},
  url={https://doi.org/10.1038/s42254-025-00813-9},
}

@INPROCEEDINGS{shor1994,
  author={Shor, P.W.},
  booktitle={Proceedings 35th Annual Symposium on Foundations of Computer Science}, 
  title={Algorithms for quantum computation: discrete logarithms and factoring}, 
  year={1994},
  volume={},
  number={},
  pages={124-134},
  doi={10.1109/SFCS.1994.365700}
}

@article{grigoryan2025quantum,
    AUTHOR = {Grigoryan, Eduard and Kumar, Sachin and Pinheiro, Placido Rogério},
    TITLE = {A Review on Models and Applications of Quantum Computing},
    JOURNAL = {Quantum Reports},
    VOLUME = {7},
    YEAR = {2025},
    NUMBER = {3},
    ARTICLE-NUMBER = {39},
    URL = {https://www.mdpi.com/2624-960X/7/3/39},
    ISSN = {2624-960X},
    DOI = {10.3390/quantum7030039}
}

@article{preskill2018nisq,
   title={Quantum Computing in the NISQ era and beyond},
   volume={2},
   ISSN={2521-327X},
   url={http://dx.doi.org/10.22331/q-2018-08-06-79},
   DOI={10.22331/q-2018-08-06-79},
   journal={Quantum},
   publisher={Verein zur Forderung des Open Access Publizierens in den Quantenwissenschaften},
   author={Preskill, John},
   year={2018},
   month=aug, 
   pages={79},
}

@inproceedings{grover1996,
    author = {Grover, Lov K.},
    title = {A fast quantum mechanical algorithm for database search},
    year = {1996},
    isbn = {0897917855},
    publisher = {Association for Computing Machinery},
    address = {New York, NY, USA},
    url = {https://doi.org/10.1145/237814.237866},
    doi = {10.1145/237814.237866},
    booktitle = {Proceedings of the Twenty-Eighth Annual ACM Symposium on Theory of Computing},
    pages = {212–219},
    numpages = {8},
    location = {Philadelphia, Pennsylvania, USA},
    series = {STOC '96}
}

@misc{eisert2025mindgapsfraughtroad,
      title={Mind the gaps: The fraught road to quantum advantage}, 
      author={Jens Eisert and John Preskill},
      year={2025},
      eprint={2510.19928},
      archivePrefix={arXiv},
      primaryClass={quant-ph},
      url={https://arxiv.org/abs/2510.19928},
      doi={10.48550/arXiv.2510.19928},
}

@article{google2019quantumsupremacy,
    title = {Quantum Supremacy using a Programmable Superconducting Processor},
    author = {Frank Arute and others},
    year = {2019},
    URL	= {https://www.nature.com/articles/s41586-019-1666-5},
    doi = {10.1038/s41586-019-1666-5},
    journal	= {Nature},
    pages = {505–510},
    volume = {574}
}

@article{ladd2010quantum,
   title={Quantum computers},
   volume={464},
   ISSN={1476-4687},
   url={http://dx.doi.org/10.1038/nature08812},
   DOI={10.1038/nature08812},
   number={7285},
   journal={Nature},
   publisher={Springer Science and Business Media LLC},
   author={Ladd, T. D. and Jelezko, F. and Laflamme, R. and Nakamura, Y. and Monroe, C. and O’Brien, J. L.},
   year={2010},
   month=mar, pages={45–53} 
}

@book{evans10,
  address = {Providence, R.I.},
  author = {Evans, Lawrence C.},
  isbn = {9780821849743 0821849743},
  publisher = {American Mathematical Society},
  refid = {465190110},
  title = {Partial differential equations},
  year = 2010
}

@article{montanaro2016quantum,
   title={Quantum algorithms: an overview},
   volume={2},
   ISSN={2056-6387},
   url={http://dx.doi.org/10.1038/npjqi.2015.23},
   DOI={10.1038/npjqi.2015.23},
   number={1},
   journal={npj Quantum Information},
   publisher={Springer Science and Business Media LLC},
   author={Montanaro, Ashley},
   year={2016},
   month=jan 
}

@article{quantinuum2025supremacy,
  title = {Computational Power of Random Quantum Circuits in Arbitrary Geometries},
  author = {DeCross, M. and others},
  journal = {Phys. Rev. X},
  volume = {15},
  issue = {2},
  pages = {021052},
  numpages = {39},
  year = {2025},
  month = {May},
  publisher = {American Physical Society},
  doi = {10.1103/PhysRevX.15.021052},
  url = {https://link.aps.org/doi/10.1103/PhysRevX.15.021052}
}

@misc{novikau2025efficientexplicit,
      title={An efficient explicit implementation of a near-optimal quantum algorithm for simulating linear dissipative differential equations}, 
      author={Ivan Novikau and Ilon Joseph},
      year={2025},
      archivePrefix={arXiv},
      primaryClass={quant-ph},
      doi={10.48550/arXiv.2501.11146}, 
}

@article{novikau2025quantumalgorithmadvection,
title = {Quantum algorithm for the advection-diffusion equation and the Koopman-von Neumann approach to nonlinear dynamical systems},
journal = {Computer Physics Communications},
volume = {309},
pages = {109498},
year = {2025},
issn = {0010-4655},
doi = {https://doi.org/10.1016/j.cpc.2025.109498},
url = {https://www.sciencedirect.com/science/article/pii/S0010465525000013},
author = {I. Novikau and I. Joseph},
}

@article{low2017optimal,
  title = {Optimal Hamiltonian Simulation by Quantum Signal Processing},
  author = {Low, Guang Hao and Chuang, Isaac L.},
  journal = {Phys. Rev. Lett.},
  volume = {118},
  issue = {1},
  pages = {010501},
  numpages = {5},
  year = {2017},
  month = {Jan},
  publisher = {American Physical Society},
  doi = {10.1103/PhysRevLett.118.010501},
  url = {https://link.aps.org/doi/10.1103/PhysRevLett.118.010501}
}

@article{low2019hamiltonian,
   title={Hamiltonian Simulation by Qubitization},
   volume={3},
   ISSN={2521-327X},
   url={http://dx.doi.org/10.22331/q-2019-07-12-163},
   DOI={10.22331/q-2019-07-12-163},
   journal={Quantum},
   publisher={Verein zur Forderung des Open Access Publizierens in den Quantenwissenschaften},
   author={Low, Guang Hao and Chuang, Isaac L.},
   year={2019},
   month={july}, pages={163} 
}

@misc{laneve2025,
      title={Generalized Quantum Signal Processing and Non-Linear Fourier Transform are equivalent}, 
      author={Lorenzo Laneve},
      year={2025},
      eprint={2503.03026},
      archivePrefix={arXiv},
      primaryClass={quant-ph},
      url={https://arxiv.org/abs/2503.03026}, 
}

@misc{silva2022,
      title={Fourier-based quantum signal processing}, 
      author={Thais de Lima Silva and Lucas Borges and Leandro Aolita},
      year={2022},
      eprint={2206.02826},
      archivePrefix={arXiv},
      primaryClass={quant-ph},
      url={https://arxiv.org/abs/2206.02826}, 
}

@misc{qsp4pde,
  author = {Hyeongjin Kim},
  title = {{qsp4pde}},
  howpublished = {\url{https://github.com/hjkqubit/qsp4pde}},
  year = {2026}
}

@article{birdsall,
  title={Particle-in-cell charged-particle simulations, plus Monte Carlo collisions with neutral atoms, PIC-MCC},
  author={Birdsall, Charles K},
  journal={IEEE Transactions on plasma science},
  volume={19},
  number={2},
  pages={65--85},
  year={2002},
  publisher={IEEE},
  doi={10.1109/27.106800}
}

@article{turner2013,
  title={Simulation benchmarks for low-pressure plasmas: Capacitive discharges},
  author={Turner, Miles M and Derzsi, Aranka and Donko, Zoltan and Eremin, Denis and Kelly, Sean J and Lafleur, Trevor and Mussenbrock, Thomas},
  journal={Physics of Plasmas},
  volume={20},
  number={1},
  year={2013},
  publisher={AIP Publishing},
  issn = {1070-664X},
  doi = {10.1063/1.4775084},
  url = {https://doi.org/10.1063/1.4775084},
}

@article{myers2021,
  title={Porting WarpX to GPU-accelerated platforms},
  author={Myers, Andrew and Almgren, A and Amorim, Ligia Diana and Bell, J and Fedeli, Luca and Ge, Lixin and Gott, Kevin and Grote, David P and Hogan, M and Huebl, Axel and others},
  journal={Parallel Computing},
  volume={108},
  pages={102833},
  year={2021},
  publisher={Elsevier},
  issn = {0167-8191},
  doi = {https://doi.org/10.1016/j.parco.2021.102833},
  url = {https://www.sciencedirect.com/science/article/pii/S0167819121000818},
}

@article{Aaronson2015,
  title = {Read the fine print},
  volume = {11},
  ISSN = {1745-2481},
  url = {http://dx.doi.org/10.1038/nphys3272},
  DOI = {10.1038/nphys3272},
  number = {4},
  journal = {Nature Physics},
  publisher = {Springer Science and Business Media LLC},
  author = {Aaronson,  Scott},
  year = {2015},
  month = Apr,
  pages = {291–293}
}

@article{camps21,
author = {Camps, Daan and Van Beeumen, Roel and Yang, Chao},
title = {Quantum {F}ourier transform revisited},
journal = {Numerical Linear Algebra with Applications},
volume = {28},
number = {1},
pages = {e2331},
doi = {https://doi.org/10.1002/nla.2331},
url = {https://onlinelibrary.wiley.com/doi/abs/10.1002/nla.2331},
eprint = {https://onlinelibrary.wiley.com/doi/pdf/10.1002/nla.2331},
year = {2021}
}
\end{document}